\documentclass[12pt]{article}
\usepackage{amsmath}
\usepackage{amsfonts}
\usepackage{amssymb}
\usepackage{latexsym}
\usepackage{pgfplots}
\usepackage{stix}
\usepackage{comment}

\usepackage{amsmath, amsfonts, amssymb,latexsym}

\input epsf.sty

\newtheorem{Theorem}{Theorem}[section]
\newtheorem{Lemma}{Lemma}[section]

\newcommand{\BEQ}{\begin{equation}}     
\newcommand{\BEA}{\begin{eqnarray}}
\newcommand{\BD}{\begin{displaymath}}
\newcommand{\EEQ}{\end{equation}}       
\newcommand{\EEA}{\end{eqnarray}}
\newcommand{\ED}{\end{displaymath}}

\newcommand{\del}{\delta}
\newcommand{\Del}{\Delta}

\newcommand{\supp}{{\mathrm{supp}}}

\newcommand{\R}{\mathbb{R}}

\renewcommand{\P}{\mathbb{P}}

\renewcommand{\L}{\mathbb{L}}
\newcommand{\Ker}{{\mathrm{Ker}}}

\def\Q{{\mathbb{Q}}}
\def\esper{{\mathbb{E}}}
\def\T{{\mathbb{T}}}

\def\Var{{\mathrm{Var}}}

\newcommand{\eop}{\hfill $\Box$}        
\newcommand{\Medskip}{\medskip\noindent}
\newcommand{\Bigskip}{\bigskip\noindent}

\renewcommand{\Im}{{\rm Im\ }}          
\newcommand{\half}{{1\over 2}}          

\newcommand{\var}{{\mathrm{var}}}

\begin{document}

\title
{\bf Exact computation of growth-rate fluctuations in random environment}

\maketitle

\vspace{2mm}
\begin{center}
{\bf J\'er\'emie Unterberger$^a$}
\end{center}

\vskip 0.5 cm
\centerline {$^a$Institut Elie Cartan,\footnote{Laboratoire 
associ\'e au CNRS UMR 7502} Universit\'e de Lorraine,} 
\centerline{ B.P. 239, 
F -- 54506 Vand{\oe}uvre-l\`es-Nancy Cedex, France}
\centerline{jeremie.unterberger@univ-lorraine.fr}

\vspace{2mm}
\begin{quote}

\renewcommand{\baselinestretch}{1.0}
\footnotesize {\bf Abstract.} 
{We consider a general class of Markovian models describing the
growth in a randomly fluctuating environment  of a clonal
biological population  having several phenotypes related by stochastic switching. Phenotypes differ e.g. by the level
of gene expression for a population of bacteria. The time-averaged growth rate of the population, $\Lambda$, is self-averaging in the limit of infinite times; it may be understood as the fitness
of the population in a context of Darwinian evolution.

The observation time $T$ being however typically finite,  the growth rate fluctuates. For $T$ finite but large, we obtain the variance of the time-averaged growth rate as the maximum of a functional based on the stationary probability distribution for the phenotypes. This
formula is general. In the case of two states, the
stationary probability was computed by Hufton, Lin and Galla    
\cite{HufLin2}, allowing for an explicit expression which
can be checked numerically.

}
\end{quote}

\vspace{4mm}
\noindent



\newpage
\tableofcontents



\section{Introduction}


\Medskip {\bf General context.} We consider in the present work a general
class of dynamical models describing the growth of several   
subpopulations    in a randomly evolving environment. The interaction between the  subpopulations  manifests
itself  only through random transitions. The 
 subpopulations at time $t$ are assumed to be large enough to take
a kinetic limit and
neglect fluctuation effects, so that the only randomness
in the dynamics comes from the environment.  The latter is
modelized by a time-continuous Markov process  $(s(t))_{t\ge 0}$
switching between a finite number of values. 

\Medskip We take our inspiration from biology, where such 'mutation'-selection models are considered \cite{KL,SkaKus,HufLin1,HufLin2} as a general framework to discuss Darwinian evolution, see also  e.g. \cite{RivWal} for a
discussion of immune strategies along the same lines.  We consider the population of a given biological species. Subpopulations are characterized by their phenotype $x$; their
differential fitnesses are accounted for by a $x$-dependent
growth-rate $k_x$. In this
context, random
transitions ('mutations') between the subpopulations are a simple model for phenotypic switching; see e.g. 
the recent book \cite{Lev} for a modern confrontation of  Lamarck's and Darwin's viewpoints. 
In general, differential gene expressions mediated by interacting networks of genes and proteins, in particular, epigenetics, account for random phenotypic changes. Known examples concern e.g.
the residual  resistance of bacteria \cite{Bal} or cancer cells \cite{Sha} to drugs, or 
the level of expression of the lactose permease of E. Coli
\cite{Cho,DekAlo}; see \cite{EldElo} for a discussion of the r\^ole of
noise in genetic circuits and more examples pertaining to
stem cell differentiation and developmental evolution. 
 An important restriction is made here by assuming that growth 
 rates $k_x$ are independent of the environment: the organisms do not "sense" the environment and mutate "blindly". Perfect sensing allows maximization of growth rate by switching at time $t$ to the state $x$ maximizing
 $(k(x|s(t))_{x\in {\cal X}}$; however, sensing comes at a cost, so actual  biological strategies are probably best described as lying somewhere in-between
 these two ideal cases. 

\Medskip Under lenient conditions (ergodicity of the phenotypic
switching generator ${\cal P}^{pheno}$, see below) the time-averaged growth-rate of the total population
$N:=\sum_x N_x$, namely,
\BEQ \frac{1}{T}\Lambda_{[0,T]}:=  \frac{1}{T}\ln(N(T)/N(0))
\label{eq:1}
\EEQ
is known to be a self-averaging quantity, namely, 
\BEQ \lambda := \lim_{T\to\infty} \frac{1}{T}\Lambda_{[0,T]}
\label{eq:2}
 \EEQ
is a constant. However, at a given, finite time-horizon $T$, 
 the variance  Var$(\frac{1}{T}\Lambda_{[0,T]})$   of $\frac{1}{T}\Lambda_{[0,T]}$ is non-zero, and scales
 like $1/T$ due to the central limit theorem. Therefore, when $T\to \infty$,   $\frac{1}{T}$ Var$(\Lambda_{0,T})$ converges to a finite,
 positive quantity which we denote (by abuse of notation)
  $\var(\lambda)$, and call {\bf asymptotic growth-rate variance}. This
 is the main quantity of interest in this article.

\Bigskip Our interest for the variance of the growth-rate arose
out of a previous work \cite{DinLacUnt1} on a much simpler, scalar
problem of maximization of capital growth rate known in the mathematical finance community as Kelly's model. The variance of the growth rate
is then interpreted as a risk; trade-offs between maximization of averaged capital growth rate and minimization of its fluctuations (measured by the variance), and
also the form of the  Pareto front associated to the
multi-objective optimization, have been discussed in details there.

\medskip\noindent  The variance of the growth rate may  presumably also be interpreted as a risk in biology, bound to influence evolution on the long term. The word 'risk', taken out of its original financial context 
into this biological context, may sound at best metaphorical. 
In finite-time protocols, however, the variance of the 
{\em relative} time-integrated growth rate is naturally interpreted as a 'risk'
for a given species in a situation  where two (or more) species
living in the same environment are competing. Let us discuss
this point specifically in a discrete-time dilution protocol. We consider two species 
$a=1,2$ growing without any interaction in the same 
environment history $(s(t))_{t\ge 0}$, with concentration
vectors ${\bf N}^a=((N^a_x)_x(t))_{t\ge 0}$. The total
concentration of species $a$ at time $t$ is $N^a(t):=\sum_x
N^a_x(t)$.  {\em The protocol (P) is the following:} at time  $T$,
we apply a global dilution factor to both species, so
that the total concentration $N(T)=N^1(T)+N^2(T)$ is back
to its initial value $N(0)$. By convention we assume that
$N(0)=1$. In the idea, the dilution could be repeated
periodically with time-period $T$; instead of a deterministic dilution, we could use a 'Fisher-Wright'
type procedure where a small number of molecules is
drawn with replacement from a large pool. Here
we content ourselves with protocol (P), which is the simplest possible case. 
Let  $\Lambda^a_{[0,T]}$ be the time-integrated growth rate
of species $a$; by definition,   the concentration at time $T$ is $ N^a(T)=N^a(0)\,  e^{\Lambda^a_{[0,T]}}.$
Let $n^i(t)=\frac{N^i(t)}{N(t)}$
be the relative concentrations.
The instantaneous growth rate is $\lambda^a(t)=\frac{d\Lambda^a_{[0,t]}}{dt}$. 
 Then
\begin{equation} \frac{dN^a}{dt}=\lambda^a(t) N^a(t), \qquad 
\frac{dN/dt}{N}=n^1 \lambda^1+n^2\lambda^2 \end{equation}
At time $T$, {\em after} dilution 
 concentrations $(N^1(T),N^2(T))$ become $(n^1(T),n^2(T))$.
Considering species $a=1$, 
\begin{equation} \frac{dn^1}{dt} =\frac{dN^1/dt}{N}- n^1 \frac{dN/dt}{N}
 = \lambda^1 n^1 - n^1 (n^1\lambda^1 + n^2\lambda^2)= (\lambda^1-\lambda^2)n^1(1-n^1) \label{eq:7}
 \end{equation}
Integrating the differential equation between $t=0$ and $t=T$,,
 \begin{equation} \int \frac{dn^1}{n^1(1-n^1)}=\ln\Big(\frac{n^1(T)/n^2(T)}{n^1(0)/n^2(0)}\Big) = \Lambda^1_{[0,T]}-\Lambda^2_{[0,T]}
 \end{equation}

 \medskip\noindent We now {\em define 
\BEQ  p(t):= \ln \Big(\frac{n^1(t)/n^2(t)}{n^1(0)/n^2(0)} \Big)
\EEQ
to be the selective advantage of  species 1  relative
to species 2}. Then
 \begin{equation} \Var(p(T))=\Var(\Lambda^1_{[0,T]}-\Lambda^2_{[0,T]}) \end{equation}

\medskip\noindent In the particular case when species 2
does not  grow, we find simply that {\em the variance
of the selective advantage of species 1 is
$\Var \Lambda^1_{[0,T]}$}, which gives a justification
for the study of this quantity, which is closely
related to the quantity called 'risk' in the case
of Kelly's model. The general case
could be studied in the same framework of this article  by considering phenotype $x$ of  species $a$ as a
'superphenotype' indexed by $(a,x)$, with no  mutation
possible between superphenotypes
$(a,x),(a',x')$ with $a\not=a'$, yielding a block-diagonal
evolution matrix; the  $\Lambda_{[0,T]}^a$,$ a=1,2$ are then interpreted as  'partial' integrated 
growth rates for each of the two blocks.

\Medskip We postpone a discussion of our results after
a more detailed mathematical presentation of our model.

\Bigskip {\bf Model.} 
We fix  two finite sets, the {\em phenotypic space} ${\cal X}$,
and the {\em environmental space} $\cal S$. Our 
 growth model  is  defined by  a linear differential equation,
\BEQ \frac{d}{dt} \vec{N}(t) = P(t) \vec{N}(t),  \label{eq:dyn}
\EEQ
in coordinates, $\frac{d}{dt} N_x(t)=\sum_{y\in {\cal X}}
P_{xy}(t) N_y(t)$, 
where $\vec{N}(t)=(N_x(t))_{x\in X}$ is a vector representing the concentrations at time $t$ of the different phenotypes $x$ of 
a given clonal population. Diagonal coefficients represent the
instantaneous growth rate of each phenotype, whereas off-diagonal coefficients represent the effect of transition rates between
phenotypes. The time-dependence of the generator of time evolution $P$ comes through a varying environment, described by a trajectory $(s(t))_{t\ge 0}$ in the  external environmental space $\cal S$.  The operators $P(t)$ can be split into the sum of two operators,

\BEQ P(t)=P^{growth}(s(t))+P^{pheno} \EEQ
where:

\textbullet\   $P^{growth}(s(t))$ is a diagonal growth matrix, with coefficients  $(P^{growth}(s(t)))_{xx}=k(x|s(t))$. Each coefficient is the difference of a replication rate and of a death rate, so
it can be positive or negative. Resulting growth rates  $k(x|s(t))$ depend on the current environmental state $s(t)$. 

\textbullet  \ $P^{pheno}=(P^{pheno}_{xy})_{x,y\in{\cal X}}$,  $  P^{pheno}_{xy}=\begin{cases}
\pi_{x|y} \qquad (x\not=y) \\ -\pi_x \qquad (x=y) \end{cases}
$
is
a Markov generator representing   random  transitions from
phenotype $y$ to phenotype $x$, with
$\pi_x:=\sum_{y\not=x} \pi_{y|x}$ inverse holding time. {\em We assume} that $P^{pheno}$ (contrary to $P^{growth}$) {\em is independent of the environment.} Also (as mentioned above), we assume the Markov
process to be irreducible, or equivalently, ergodic. Note that the fact that the sum
$\sum_{x\in{\cal X}} P^{pheno}_{xy}$ of coefficients on a
column vanishes is equivalent to the conservation of the
total concentration $\sum_{x\in{\cal X}} N_x$ when $k=0$.

\Medskip In the case e.g. of two states $(|{\cal X}|=|{\cal S}|=2)$, $P(s(t))=P_1$ or $P_2$ depending
on whether $s(t)=1$ or $2$, and $P_1=\left(\begin{array}{cc}
 k(1|1)-\pi_{1} & \pi_{1|2}  \\ \pi_{2|1} & k(2|1)-\pi_{2} \end{array}\right), P_2=\left(\begin{array}{cc} k(1|2)-\pi_{1} & \pi_{1|2}  \\ \pi_{2|1} & k(2[2)-\pi_{2} \end{array}\right)$.

\Bigskip  The time-evolution $(s(t))_{t\ge 0}$ of the environment is given
by an irreducible continuous-time Markov process,

\BEQ s'\overset{\kappa_{s|s'}}{\to} s, \qquad s'\not=s\in{\cal S} \label{eq:sbars}
\EEQ  
with rates $(\kappa_{s|s'})_{s\not=s'}$.   
The transition rate out of state $s'$ (inverse holding
time) is $\kappa_{s'}:=\sum_{s\not=s'} \kappa_{s|s'}$.

\Bigskip Let $N(t):=\sum_x N_x(t)$ be the total
population. For simplicity, we assume that the population
at time $0$ is normalized, i.e. $N(0)=1$.  The Lyapunov
exponent
characterizing the system in the long-time limit is  the
average growth rate of the
population, namely,
$\lambda:=\lim_{T\to\infty} 
\frac{1}{T} \Lambda_{[0,T]}$, where  $\Lambda_{[0,T]}:=
\log(N(t))$, see eqs. (\ref{eq:1}), (\ref{eq:2}).

\Medskip  
 In the limit when the time-scale of environmental transitions
 is much larger than the time-scale of phenotypic transitions (see
 Kussell-Leibler \cite{KL}), the
 system aligns most of the time in the direction corresponding to
 the highest eigenvalue of $P(t)$, and it can be argued that $\lambda$ is largest
when phenotypic switching rates follow closely environmental
transition rates, implying in particular a {\em 
bet-hedging strategy} for the population, i.e. the coexistence of different phenotypes at all times, including those not maximizing the instantaneous growth rate.  In this limit, both 
$\lambda$ and var$(\lambda)$  \cite{DinLacUnt2} may be computed. However, they cannot in general. See \cite{HufLin1,HufLin2} for a detailed discussion of the two-state model $(|{\cal X}|=|{\cal S}|=2)$, and 
\cite{HufLin1,HufLin2,RivWal} for phase diagrams also involving other strategies.

\Medskip Formally, the solution of (\ref{eq:dyn})  may be written
 as a time-ordered integral,\\  $[A](t)=\overrightarrow{\exp}\Big(\int_0^t dt'\, 
P(t')\Big) [A](0)$. Except in very simple cases, however, the integral cannot be computed. Discretizing time, one obtains instead
products of random matrices. The long-time limit is deterministic
due to self-averaging, but not obtained as the result of an explicit computation. One may also try to solve for
the joint probability distribution ${\cal P}_t(\phi,s)$ by means of a master equation, where
$s\in {\cal S}$ and $\phi:=(\frac{N_x(t)}{N(t)})_{x\in {\cal X}}$ is a vector giving the proportion of each phenotype in the population. Since the trajectory $(\phi(t))$ between two
 successive environment jump times is deterministic, this
 gives rise to a {\em piecewise deterministic  Markov
 process} (PDMP for short). It is  proved in Hufton-Lin-Galla-McKane \cite{HufLin1,HufLin2} 
that ${\cal P}_t(\phi,s)$ converges when $t\to\infty$ to
a stationary distribution ${\cal P}_{stat}(\phi,s)$, from
which one deduces the asymptotic environment-dependent density
$\rho_s(x):=\int \phi_x \, {\cal P}_{stat}(\phi,s) \, d\phi$. 
One then concludes to the existence when 
$T\to\infty$ of 
a {\em time-averaged density} $\bar{\rho}$ -- equal by
self-averaging to the  average $\lim_{T\to\infty} \Q_{[0,T]}[\rho^e(x)]$ of the empirical density $\rho^e(x):=\frac{1}{T}
\int_0^T dt\, \del_{x(t),x}$
w.r. to all environmental {\em trajectories} --, which identifies by the ergodic theorem with
the average of $\rho_s$ over  environmental  states, 

\BEQ \bar{\rho}(x):=\lim_{T\to\infty}  \frac{1}{T} \int_0^T dt\, \rho_t(x)= \sum_s {\cal Q}_s \rho_s(x).  \label{eq:rhobar}
\EEQ
See next paragraph for notations. Unfortunately, ${\cal P}_{stat}$ can be computed in closed form only in the case of two states.

\Bigskip {\em Further hypotheses and notations.} We assume
that the environmental Markov chain satisfies
local balance, namely,
the stationary measure ${\cal Q}$ for the {\em environmental Markov
chain} (\ref{eq:sbars}) satisfies $\kappa(s|s'){\cal Q}(s')=\kappa(s'|s){\cal Q}(s)$.
We further denote $\Q_{[0,T]}$  the law
of environmental trajectories $s:[0,T]\to\R$ under the stationary
environmental Markov chain, and by  $\Q_{[0,T]}[\cdots]$ or (for short) $\langle \cdots\rangle$ the
 expectation with respect to $\Q_{[0,T]}$.

\Bigskip  {\bf  Results of the article.} 
Our results concern the {\em asymptotic growth-rate variance}, by definition,
\BEQ {\mathrm{var}}(\lambda)=\lim_{T\to\infty}  \frac{1}{T}   \Big( \langle (\Lambda_{[0,T]})^2\rangle - \langle \Lambda_{[0,T]}\rangle^2  \Big).  \label{eq:varlambda}
\EEQ

\Medskip  Our first result, holding for an arbitrary
number of phenotypes and environmental states (see Theorem
\ref{th:main2}), yields
var$(\lambda)$ as the solution of a variational formula
obtained in terms of the stationary PDMP probability
measure $\P$. 

\Medskip   This result is somewhat abstract since
$\P$ is not known in general. However,  in the case when
there are only two phenotypic and two environmental
states ($|{\cal X}|=|{\cal S}|=2$), $\P$ is known
from the work of Hufton, Lin, Galla and McKane
(\cite{HufLin1,HufLin2}), which allows an explicit
computation of the variance in terms of a double integral, see
Theorem \ref{th:final}.
 Our result has been checked
numerically by L. Dinis and D. Lacoste, and is used
in our work in preparation \cite{DinLacUnt2} to explore
the Pareto front featuring the mean and the variance of the time-averaged growth
rate, in close analogy with \cite{DinLacUnt1}.

\Bigskip {\em Plan of the article.}  Our general
results are presented in Section 2. The explicit
computation in the particular case
 ($|{\cal X}|=|{\cal S}|=2$) is given in Section 3.


\section{Stationary distribution and variational formula
for the growth-rate variance}


{\bf Hufton-Lin's PDMP reformulation.} The state of
the system at time $t$ is characterized by (i) the total
concentration $N(t):=\sum_{x\in {\cal X}} N_x(t)$ and by the relative
fractions
$\phi_x(t):=\frac{ N_x(t)}{N(t)}$; (ii) the environmental state $s(t)$.  Rewriting (\ref{eq:dyn}) in terms of these new variables, one gets  (see  \cite{SkaKus} or \cite{HufLin1})

\BEQ \frac{d(\log N)}{dt}=\frac{dN/dt}{N}=\sum_x k(x|s)\phi_x
\label{eq:dlogN/dt}
\EEQ

\BEQ \frac{d\phi_x}{dt}=\frac{1}{N}\frac{d N_x}{dt} - 
\phi_x \frac{d(\log N)}{dt}= v_x(\phi|s)
\EEQ

with
\BEQ v_x(\phi|s):=\Big\{ k(x|s) (1-\phi_x) - \sum_{y\not=x} 
k(y|s)\phi_y\Big\} \, \phi_x \ -\pi_x \phi_x + \sum_{y\not=x}
\pi(x|y)\phi_y.
\EEQ

Since  $\log(N(t))=\int_0^t dt'\,  \sum_x k(x|s(t'))\phi_x(t')$, 
time-trajectories $(\phi(t);s(t))_{t\ge 0}$ suffice to determine the time concentrations. Note that $\sum_s \phi_s=1$ is 
a conserved quantity.

\Medskip
The time-evolution of the coupled system 
$\Phi:=(\phi;s)$ is a so-called PDMP (piecewise deterministic
Markov process), a particular type of Feller Markov process
(see \cite{RevYor}, Chap. VII for an introduction) with generator
$\mathbb L$ acting on  a space $C_0(\R^{X}\times {\cal S})$  identified with the space of continuous vector-valued
$\mathbb L$ functions $\{f\equiv (f_s)_{s\in {\cal S}}, f_s\in C_0(\R^{X})$ with components indexed by $\cal S$, 
\BEQ ({\mathbb L} f)_s(\phi)=\sum_{s'} {\mathbb L}_{s,s'}(\phi)
f_{s'}, \qquad {\mathbb L}_{s,s'}(\phi)=\begin{cases}
v(\phi|s)\cdot\nabla - \kappa_s \qquad s=s' \\ 
\kappa_{s'|s} \qquad s\not=s' \end{cases}
\EEQ

with $v(\phi|s)\cdot\nabla=\sum_{x\in{\cal X}} v_x(\phi|s) \partial_{\phi_x}$
"convection term" in the space $\R^{{\cal X}}$. 
Because ${\mathbb L}={\cal L}_{\cal S} + $ diag$(v(\cdot|s)\cdot\nabla)$, where ${\cal L}_{\cal S}$ is the generator
 of the environmental Markov chain, it is immediately checked that ${\mathbb L}({\mathbb 1})=0$,
 where ${\mathbb 1}=(1,\ldots,1)^t$ is the constant function. 
 
\Medskip Because  the environmental Markov chain is
 irreducible, it can be proved that 
the above PDMP has a unique stationary probability measure, $\P=\P(\phi;s)$. Letting for short $\P_s(\phi)=\P(\phi;s)$, the
normalization condition is $\sum_s\int d\phi\, \P_s(\phi)=1$.
By definition, $\P$ generates the kernel of the adjoint operator
${\mathbb L}^*$ in $L^2(\R^{X}\times {\cal S})$:  letting
\BEQ ({\mathbb L}^* f)_s(\phi)=\sum_{s'} {\mathbb L}^*_{s,s'}
f_{s'}, \qquad {\mathbb L}^*_{s,s'}=\begin{cases}
-\nabla\cdot (v(\cdot|s)\, \cdot) - \kappa_s \qquad s=s' \\ 
\kappa_{s|s'} \qquad s\not=s' \end{cases}  \label{eq:L*}
\EEQ
featuring the adjoint operator $-\nabla\cdot v(\phi|s):=
\Big(v(\phi|s)\cdot\nabla\Big)^*=-\sum_x \partial_{\phi_x}(v_x(\phi|s) \ \cdot)$,  
one has ${\mathbb L}^* (\P)=0$.  {\em Warning:} the dot
inside the parenthesis after the drift velocity $v$ 
emphasizes that $\nabla\cdot (v(\cdot|s)\, \cdot)$ or
$\partial_{\phi_x}(v_x(\phi|s) \ \cdot)$ is understood as
an operator, i.e. it acts on a function component $f_s$ as
$\nabla\cdot (v(\cdot|s)\, f_s(\phi))$ or $\partial_{\phi_x}(v_x(\phi|s) \ f_s(\phi))$.
 
\Medskip The natural $L^2$ space in this problem is not 
$L^2(\R^{|X|}\times {\cal S})$ with its standard scalar product
\BEQ (f,g):=\sum_s \int d\phi\, f_s(\phi)g_s(\phi), \EEQ
but the $\mathbb P$-weighted space $L^2({\mathbb P})$ with
scalar product
\BEQ (f,g)_{\mathbb P} :=\sum_s \int d\phi\, {\mathbb P}_s(\phi) f_s(\phi)g_s(\phi). \EEQ
The adjoint  of $\L$ w. r. to the latter scalar product
will be denoted $\L^{\dagger}$; since $({\mathbb L}f,g)_{\mathbb P}=(\P g, {\mathbb L}f)=(\P^{-1} {\mathbb L}^* \P g,f)_{\P}$, one has 
\BEQ {\mathbb L}^{\dagger}=\P^{-1} {\mathbb L}^* \P. \EEQ
Note that $\Ker {\mathbb L}^{\dagger}$ is generated by 
$\mathbb 1$. 

\Medskip For the sequel we also need to introduce the
{\em symmetrized generator},
\BEQ \L_{sym}:=\half (\L + \L^{\dagger}).\EEQ
By definition, $\L_{sym}=\half \P^{-1} (\P \L + \L^* \P)$. 
Explicit computation yields for diagonal coefficients $-\half(\P\L + \L^* \P)_{s,s}=\half
\Big\{ \P_s (-v(\cdot|s)\cdot\nabla + \kappa_s) + ( 
\nabla\cdot v(\cdot|s)+\kappa_s)\P_s\Big\}=\half \sum_x \partial_{\phi_x} (v_x(\phi|s)\P_s) + \kappa_s \P_s= \half (\kappa_s \P_s + 
\sum_{s'\not=s} \kappa_{s|s'}\P_{s'})$ (in the last equality we
have used the stationarity of $\P$), and for off-diagonal coefficients, $-\half (\P\L+\L^*\P)_{s,s'}=-\half (\kappa_{s|s'} \P_{s'} + \kappa_{s'|s} \P_s)$  ($s\not=s'$).  
Taking scalar products w.r. to the $\P$-weighted scalar product compensates the extra $\P^{-1}$ weight in front of $\L_{sym}$, and we then get an explicit expression of the non-negative
(degenerate) quadratic form associated to $-\L_{sym}$, 

\begin{Lemma}[quadratic form associated to $-\L_{sym}$]
\label{lem:Lsym}
\BEQ  (-\L_{sym}f,f)_{\P}=\half \int d\phi\ \sum_{s\not=s'} 
\kappa_{s|s'} \P_{s'}(\phi) \, (f_s(\phi)-f_{s'}(\phi))^2.
\EEQ 
\end{Lemma}

Note that ${\mathbb 1}\in\Ker \L_{sym}$, as was the case
for $\L$ and $\L^{\dagger}$, but 
\BEQ \Ker\L_{sym}=\{k(\phi) {\mathbb 1}, \qquad k:\R^{\cal X}\to\R\}
\EEQ
 is infinite-dimensional. This is due to the fact that
 symmetrizing has killed the differential part connecting
 the different values of $\phi$.
Fixing $\phi$, one has: $\Ker \L_{sym}(\phi)=\R 1$ with
$1=(1 \cdots 1)^t\in\R^{\cal X}$, 
and $\Im \L_{sym}(\phi)=1^{\perp}(\phi):=\{k\in \R^{\cal X} \ |\ 
\sum_x \P_s(\phi) k_s=0\}$. 

\Bigskip {\bf Average growth rate.} Let $\tilde{\esper}$ be the average w. r. to the measure $\tilde{\P}$ of 
the trajectories of the stationary PDMP. Start from (\ref{eq:dlogN/dt}),  and time-integrate  between $0$ and $T$. 
  Since the process
is asymptotically stationary, one gets when $T\to\infty$
\BEQ \lambda=(\P,f_{growth})=\sum_s \int d\phi\, \P_s(\phi) f_{growth}(\phi;s) \EEQ
where
\BEQ f_{growth}(\phi;s)=\sum_x k(x|s)\phi_x \EEQ
is the {\em growth functional}.

\Bigskip {\bf Variance of the growth rate.} By definition,
the variance of the integrated growth rate is equal (up to
normalization) to the
variance of the integrated growth-rate function $\int_0^T dt\,
 f_{growth}(\Phi(t))$, so that
\BEQ {\mathrm{var}}(\lambda)=\lim_{T\to\infty} \frac{1}{T}
\tilde{\mathbb V}\Big( \int_0^T dt\, f_{growth}(\Phi(t)) \Big)
\EEQ
with $\tilde{\mathbb V}(\cdot)$=variance w.r. to the measure $\tilde{\P}$ of the trajectories of the stationary PDMP.

\begin{Theorem}[preliminary formula for the variance]  \label{th:main1}
Let $\del f_{growth}:=f_{growth}-\lambda$, then
\BEQ  {\mathrm{var}}(\lambda)=2 ((-\L)^{-1} \del f_{growth},\del f_{growth})_{\P}.
\EEQ
\end{Theorem}

\Medskip {\bf Proof.} The result is standard and completely general: 
it states that $\frac{1}{T}\tilde{\mathbb V}\Big( \int_0^T dt\, f(\Phi_t)\Big)\to_{\T\to\infty} 2 ((-\L)^{-1} \del f,\del f)_{\P}$
if $\L$ is a Feller generator, $f$ is a $C_0$ function, and
$\del f:=f-\esper[f]$, where $\esper[\cdot]$ is the expectation
w.r. to the stationary probabiliy measure. Namely (see e.g.
\cite{CazHar}) $(-\L)^{-1}=\int_0^{+\infty} d\theta\,  e^{\theta\L}$ is defined on the subspace of $C_0$ functions with zero
average, which contains in particular $\del f$. Now, letting
$\esper[\cdot]:=(\P,\cdot)$ be the $\P$-average,
\BEA && 
\frac{1}{T}\tilde{\mathbb V}\Big( \int_0^T dt\, f(\Phi_t)\Big) 
=
\frac{2}{T} \int_0^T dt\, \int_0^t dt'\, \tilde{\esper}[\del f(\Phi_t)
\del f(\Phi_{t'})] \nonumber\\
&&\sim_{T\to\infty} \frac{2}{T} \int_0^T dt \int_0^t d\theta\, \esper[ \del f  \ \ (e^{\theta\L}\del f)]  \qquad 
{\mathrm{by\ asymptotic\ stationarity}} \nonumber\\
&&= \frac{2}{T}  \int_0^T dt  \, \esper[ \del f \ \  (-\L)^{-1}(1-e^{t\L}) \del f]
\nonumber\\
&&\to_{T\to\infty}  2 ((-\L)^{-1} \del f,\del f)_{\P}.
\EEA  

\hfill \eop

\Bigskip  We may now prove our main formula, in the form
of a Legendre transform,

\begin{Theorem} [Variational formula for the variance] \label{th:main2}
Let $f$ be a $C_0$-function with 0 average, i.e. $(\P,f)=0$. Then
\BEQ \half ((-\L)^{-1} f,f)_{\P} = 
\sup_a  \Big\{ (f,a)_{\P} - \half ((-{\L}_{sym})^{-1} 
\L^{\dagger} a,\L^{\dagger} a )_{\P} \Big\}.  
\label{eq:th2}
\EEQ

\end{Theorem}

\noindent The supremum in the formula is over the set of all 
$C_0$-functions $a$, with the convention that 
$((-{\L}_{sym})^{-1} 
\L^{\dagger} a,\L^{\dagger} a )_{\P}\equiv +\infty$ if 
$\L^{\dagger}a\not\in \Im (\L_{sym})$, meaning that we can restrict to the subspace of functions $a$ such that 
\BEQ \L^{\dagger}a \in \Im (\L_{sym}).
\EEQ
 Since $(f,{\mathbb 1})_{\P}=0$ and 
$\L^{\dagger} {\mathbb 1}=0$, one can further restrict to
the  hyperplane ${\mathbb 1}^{\perp}:=\{f\ |\ (f,\mathbb 1)_{\P}=0\}$ of zero average functions. 

\smallskip \noindent The Theorem is used in the sequel with $f=\del f_{growth}$.

\Medskip
{\bf Proof.} Since $f\in {\mathbb 1}^{\perp}$, $(-\L)^{-1} f$
is well-defined. Let $g:=(-\L)^{-1}f$, then 
\BEA  \half ((-\L)^{-1} f,f)_{\P} &=&  \half(g,(-\L)g)_{\P} \nonumber\\
&=& \half  (g,(-\L_{sym})g)_{\P}.  \label{eq:29}
\EEA

\Medskip We now argue that (though the matrix $-\L_{sym}(\phi)$
is not one-to-one) the scalar product $(-\L_{sym}^{-1}h,h)_{\P}$ 
may be defined unambiguously when $h\in \Im(\L_{sym})$. Namely,
$\Ker \L_{sym}(\phi)=\R\,  1$, so $(-\L_{sym})^{-1} h$ is determined only up to the addition of $c(\phi)1$, where $c(\phi)$ is some
scalar function. However, since $\Im \L_{sym}(\phi)=1^{\perp}(\phi)$, the scalar product $(-\L_{sym}^{-1}h,h)_{\P}$ is
independent of the choice of the function $c$.

\Medskip 
The  expression in (\ref{eq:29}) is equal to $\sup_h \Big\{ (g,h)_{\P}
- \half ( (-\L_{sym})^{-1}h,h)_{\P} \Big\}$, with the same
convention, namely, $((-{\L}_{sym})^{-1} 
h,h)_{\P}\equiv +\infty$ if 
$h\not\in \Im (\L_{sym})$. Namely, the functional ${\cal F}: h
\mapsto  (g,h)_{\P}
- \half ( (-\L_{sym})^{-1}h,h)_{\P} $ is concave, and attains its maximum at $h$ satisfying the extremum equation $\frac{\del {\cal F}}{\del h}=0$, namely, $(-\L_{sym}^{-1})h=g$, or $h=-\L_{sym} g$.  

\Medskip Since $\Im \L^{\dagger}=(\Ker \L)^{\perp}={\mathbb 1}^{\perp}  \supset \Im \L_{sym}$, one may replace $h$ by $-\L^{\dagger} a$, with $a=-(\L^{\dagger})^{-1}h$. Then $(g,h)_{\P}= ((-\L)^{-1} f, (-\L)^{\dagger} a)_{\P}=(f,a)_{\P}$, 
yielding (\ref{eq:th2}).

\hfill \eop

\Bigskip {\em Characterization of the subspace for
optimization.} Let 
\BEQ U:=\{a \ |\ \L^{\dagger}a\in \Im (\L_{sym})\}. 
\label{eq:U}
\EEQ
 As already mentioned, $\L^{\dagger} a$ belongs to $\Im(\L_{sym})$ if and only if $\forall \phi, \sum_s  \P_s(\phi) (
\, \L^{\dagger} a)(\phi;s)=0$. Recalling that $\L^{\dagger}=
\P^{-1} \L^* \P$, this is equivalent to the condition
\BEQ \forall \phi, \  \sum_s \L^*(\P(\phi) a(\phi))(s)=0.\EEQ
Now, by (\ref{eq:L*}),\\  $\sum_s \Big(-\L^*(\P(\phi) a(\phi))\Big)_s=\sum_s
\Big(\nabla\cdot 
(v(\phi|s) \P_s(\phi)a_s(\phi)) + \kappa_s \P_s(\phi) a_s(\phi)
\Big) - \sum_{s,s'} \kappa_{s|s'} \P_{s'}(\phi) a_{s'}(\phi)
$. Since $\sum_s \kappa_{s|s'}=\kappa_{s'}$, there remains
only a "divergence" term (sum over environmental states of
weighted divergence of the product $v(\cdot|s)a_s(\cdot)$):
\BEQ \Big(a\in U\Big) \Leftrightarrow \Big(
\forall \phi, \sum_s \nabla\cdot 
(v(\phi|s) \P_s(\phi)a_s(\phi))=0 \Big)
\EEQ


\section{Explicit formula in the two-state model}


We assume here that ${\cal X}=\{1,2\}$ and ${\cal S}=\{1,2\}$.
Since $\phi_1+\phi_2=1$, there is only one free
variable, which we choose to be $\phi=\phi_1$, and write for short $\partial=\partial_{\phi}$.  Notations in \cite{HufLin2}
are similar to ours, with $\mu_s^A\equiv k(A|s)$, $(p,q)\equiv 
(\pi_1,\pi_2)$,  $\lambda\equiv \kappa$, $\Pi(\phi,s)\equiv \P_s(\phi)$, and
states and environments indexed by $0,1$ instead of $1,2$.
Let $\Del_s:=k(1|s)-k(2|s)$, $s=1,2$. {\em We assume in the sequel
that $\Del_1>0>\Del_2$: phenotype $s$ grows fastest in 
environment $s$.}   Then (particularizing
the formulas found in the previous section)
\BEQ \begin{cases} v_1(\phi)=\Del_1 \phi(1-\phi)-\pi_1 \phi + \pi_2 (1-\phi) \\ v_2(\phi)=\Del_2 \phi(1-\phi)-\pi_1 \phi + \pi_2 (1-\phi) \end{cases}  \label{eq:v-2-states}
\EEQ

\BEQ \L=\left[\begin{array}{cc} v_1\partial - \kappa_1 & 
\kappa_1 \\ \kappa_2 & v_2\partial - \kappa_2 \end{array}\right],
\qquad \L^*= \left[\begin{array}{cc} -\partial v_1 - \kappa_ 1 
& \kappa_2 \\ \kappa_1 & -\partial v_2-\kappa_2 
\end{array}\right]
\EEQ
 
\BEA  -\L^{\dagger} & = & \left(\begin{array}{cc} \P_1^{-1} & \\
& P_2^{-1} \end{array}\right)    \left(\begin{array}{cc} 
(\partial v_1+\kappa_1)\P_1 & -\kappa_2 \P_2 \\
-\kappa_1 \P_1 & (\partial v_2 +\kappa_2)\P_2 \end{array}\right)
\nonumber\\
 & = & \left(\begin{array}{cc} \P_1^{-1} & \\
& \P_2^{-1} \end{array}\right) 
 \left(\begin{array}{cc}  v_1 \P_1 \partial + \kappa_2 \P_2 
 & -\kappa_2 \P_2 \\ -\kappa_1 \P_1 & v_2 \P_2 \partial +
 \kappa_1 \P_1 \end{array}\right)
\EEA 

The stationarity equation $\L^* (\P)=0$ is equivalent to
\BEQ -(v_2 \P_2)'=(v_1 \P_1)'=-\kappa_1 \P_1+\kappa_2 \P_2
\label{eq:stat-eq-2-states}
\EEQ
As proved in \cite{HufLin2}, the stationary probability
$\P$ actually satisfies $v_1\P_1+v_2\P_2=0$; substituting
for $\P_2$ in the stationarity equation yields a one-dimensional transport equation which can be be solved explicitly. First,
\BEQ \supp(\P_1)=\supp(\P_2)=(\phi_2^+,\phi_1^+),
\EEQ
 where $\phi^+_{1}$, resp. $\phi^+_2$, is the largest, resp. smallest solution of the quadratic 
equation $v_1(\phi)=0$, resp. $v_2(\phi)=0$; they correspond to 
the stable fixed point of the characteristic equation in 
environment $s$.  We let
$\phi^-_s$ be the  second solutions of $v_s(\phi)=0$, $s=1,2$. Explicitly,
\BEA  \begin{cases}  \phi_1^{\pm}=\frac{\Del_1 - (\pi_1+\pi_2) \pm \sqrt{(\Del_1-(\pi_1+\pi_2))^2 + 4\Del_1 \pi_2}}{2\Del_1}, 
\nonumber\\  \phi_2^{\pm}=\frac{|\Del_2| + (\pi_1+\pi_2) \mp \sqrt{(|\Del_2|+(\pi_1+\pi_2))^2 - 4|\Del_2| \pi_2}}{2|\Del_2|}.
\end{cases} 
\EEA
\BEQ  \label{eq:phi1phi2} \EEQ
 
Diagonalizing the generator $P_s=P^{growth}(s)+P^{pheno}$, $s=1,2$, one also 
seees that eigenvalues are $k_1\phi_1^{\pm}$ for $s=1$, and
$k_2(1-\phi_2^{\pm})$ for $s=2$.   Note that  
\BEQ \phi_1^-<0<\phi_2^+ <\phi_1^+ <1< \phi_2^-. \EEQ
Now,

\BEQ  \P_1(\phi)= \frac{{\cal N}_1}{\Del_1} (\phi_1^+-\phi)^{g-1}
(\phi-\phi_2^+)^h f_1(\phi) \label{eq:2-states-P1} \EEQ
\BEQ \P_2(\phi)=  \frac{{\cal N}_2}{|\Del_2|} (\phi-\phi_2^+)^{h-1}
 (\phi_1^+ - \phi)^g 
 f_2(\phi)  \label{eq:2-states-P2}
\EEQ
with 
\BEQ 
g=\frac{\kappa_1}{\Del_1(\phi_1^+-\phi_1^-)}, \qquad h=\frac{\kappa_2}{|\Del_2| (\phi_2^- -\phi_2^+)} 
\EEQ
positive constants, and 
\BEQ f_1(\phi)=(\phi-\phi_1^-)^{-g-1} (\phi_2^--\phi)^{-h}, \EEQ
\BEQ f_2(\phi)=(\phi-\phi_1^-)^{-g} (\phi_2^- - \phi)^{-h-1}
\EEQ
Normalization constants ${\cal N}_1$, ${\cal N}_2$ ensure
that $\int_0^1 d\phi\, \P_s(\phi)={\cal Q}_s=
\begin{cases} \frac{\kappa_2}{\kappa_1+\kappa_2} \qquad s=1 
\\ \frac{\kappa_1}{\kappa_1+\kappa_2} \qquad s=2 \end{cases} $. Formulas
(10a),(10b) in \cite{HufLin2} feature only one normalization
constant ${\cal N}={\cal N}_1,{\cal N}_2$. It actually
follows from  the vanishing of the probability
current that ${\cal N}_1={\cal N}_2$ (indeed, evaluating (26) in \cite{HufLin1} at $\phi=\phi_2^+$ yields
 $\int_{\phi_2^+}^{\phi_1^+} d\phi_,  (\kappa_1 \P_1-\kappa_2 \P_2)(\phi)=0$). 

\Medskip  Note that
$\P_1(\phi)$ vanishes at the left end of the support $(\phi_2^+)$; on the other hand, $\P_1$  diverges like $(\phi_1^+-\phi)^{g-1}$  at the right end
of the support $(\phi_1^+)$ if $g<1$. But $v_1$ vanishes
to first order at $\phi_1^+$, so  $(v_1\P_1)(\phi)\sim_{\phi\to\phi_1^+} c(\phi_1^+ - \phi)^g\to 0$.

\Bigskip
{\centerline{
\begin{tikzpicture}[scale=6]
\draw[ultra thick](0,-0.1)--(0,0.5);
\draw[ultra thick](1,-0.1)--(1,0.5);
\draw[red](-0.15,0)--(0.15,0);\draw(0.15,0)--(0.85,0);
\draw[red](0.85,0)--(1.15,0);
\draw(0.03,-0.05) node {$0$};
\draw(0.97,-0.05) node {$1$};
\draw[red](0.15,0) node {$\times$};
\draw[red](0.85,0) node {$\times$};
\draw[red] (0.15,-0.05) node {$\phi_2^+$};
\draw[red] (0.85,-0.05) node {$\phi_1^+$};
\draw[dashed](0.85,0)--(0.85,1);
\draw[red](-0.15,0) node {$\times$};
\draw[red](-0.15,-0.05) node {$\phi_1^-$};
\draw[red](1.15,-0.05) node {$\phi_2^-$};
\draw[red](1.15,0) node {$\times$};
\draw[samples=100,domain=0.15:0.8] plot(\x, { 0.06 * pow(\x+0.15,-1.1)  *pow(\x-0.15,0.2)*
pow(0.85-\x,-0.9) * pow(1.15-\x,-0.2)   }  ); 
\end{tikzpicture}
}}

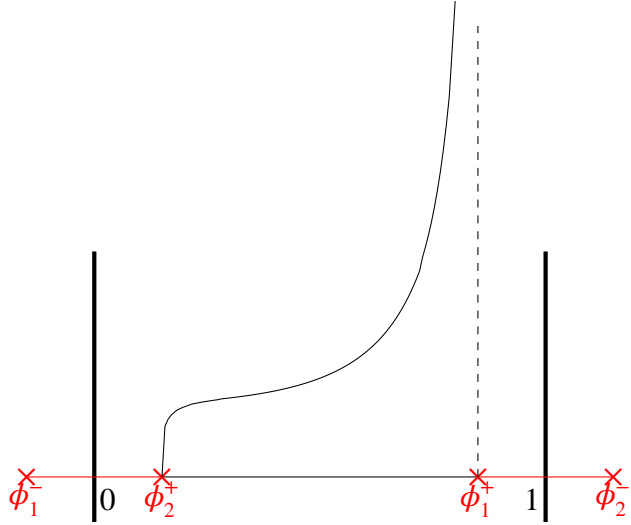
\Medskip {\bf\small Plot of $\P_1(\phi)$ $(g<1)$.   Parameters: $|\phi_1^-|=\phi_2^+=1-\phi_1^+=|1-\phi_2^-|=0.15$, $h=0.2$, $g=0.1$.} 

\Bigskip
Then,

\BEA && \P (-\L_{sym})=-\half \P (\L + \L^{\dagger}) \nonumber\\
&& =\half \left[
\begin{array}{cc} \P_1(-v_1\partial +\kappa_1) & -\kappa_1 \P_1
\\ -\kappa_2 \P_2 & \P_2 (-v_2\partial+\kappa_2) 
\end{array}
\right] + \half \left[
\begin{array}{cc} v_1 \P_1\partial+\kappa_2 \P_2 & -\kappa_2 \P_2
\\ -\kappa_1 \P_1 &   v_2\P_2 \partial+\kappa_1 \P_1
\end{array}
\right] \nonumber\\
&& = \half (\kappa_1 \P_1 + \kappa_2 \P_2) \, \times\, 
\left[\begin{array}{cc} 1 & -1 \\ -1 & 1 \end{array}\right]
\EEA
whence
\BEQ  (-\L_{sym}f,f)_{\P}=\half \int d\phi\, (\kappa_1 \P_1(\phi) + \kappa_2 \P_2(\phi))(f_1(\phi)-f_2(\phi))^2  \label{eq:Lsym22}
\EEQ
as a particular case of Lemma \ref{lem:Lsym}.

\Medskip Note also that the kernel of $-\L_{sym}(\phi)$ is (as expected)
$\R(1 \, ,\,  1)^t$ for any fixed $\phi$.  Its lines $(L_1),(L_2)$ are proportional with $(L_1)=-\frac{\P_1^{-1}}{\P_2^{-1}} (L_2)= \frac{\P_2}{-\P_1} \, (L_2)$,
so  
\BEQ \Im (-\L_{sym})=\R \left(\begin{array}{cc} \P_2(\phi) &
 -\P_1(\phi) \end{array}\right)
\EEQ
for
any fixed $\phi$. If $f\in \Im \L_{sym}$, then 
$(-\L_{sym})^{-1} f= (\kappa_1 \P_1+\kappa_2 \P_2)^{-1}
 \left[\begin{array}{c} \P_1 f_1 \\ \P_2 f_2 \end{array}\right]$
 modulo $\phi$-dependent vectors in the direction $\left[
 \begin{array}{c} 1 \\ 1\end{array}\right]$.  (Namely, 
 letting $f\in \Im (-\L_{sym})$ so that $\P_1 f_1=-\P_2 f_2$, 
 $\P (-\L_{sym}) (\kappa_1 \P_1 + \kappa_2 \P_2)^{-1}
 \left(\begin{array}{c} \P_1 f_1 \\  \P_2 f_2 \end{array}\right)
 = \half \left(\begin{array}{cc} 1 & -1 \\ -1 & 1 \end{array}
 \right) \left(\begin{array}{c} \P_1 f_1 \\ \P_2 f_2
 \end{array}\right) = \P \left(\begin{array}{c} f_1 \\ f_2
 \end{array}\right)$.)

\Bigskip {\em Characterization of the subspace for
optimization.} Recall from (\ref{eq:U}) that $U=\{a \ |\ \L^{\dagger}a\in \Im (\L_{sym})\}$.  The divergence operator $\nabla\cdot$ 
is one-dimensional here, yielding the equation
$\partial(v_1 \P_1 a_1+v_2\P_2 a_2)=0$.
We thus have $v_1\P_1 a_1=-v_2 \P_2 a_2+c$ for some constant $c$ , i.e., assuming that $c=0$ (see below) 
\BEQ a_1=a_2.
\EEQ
Then, if $a_1=a_2$, 
\BEQ -\L^{\dagger}a=\P^{-1} \left(\begin{array}{c} v_1 \P_1 
a'_1  \\ v_2 \P_2 a'_2 \end{array}\right)= \left(\begin{array}{c} v_1 a'_1 \\ v_2 a'_1 \end{array}\right)
\EEQ
so
\BEA && \half ((-\L_{sym})^{-1} \L^{\dagger} a, \L^{\dagger} a)_{\P}= \half 
\Big(  \left(\begin{array}{c} \frac{\P_1 v_1  a'_1}{\kappa_1
\P_1 +\kappa_2 \P_2} \\  \frac{\P_2 v_2 a'_1}{\kappa_1
\P_1 +\kappa_2 \P_2} \end{array}\right), \left(
\begin{array}{c} v_1 a'_1 \\ v_2 a'_1 \end{array}\right) \Big)_{\P}  \nonumber\\
&&= \int_{\phi_2^+}^{\phi_1^+} d\phi\, \frac{ (v_1(\phi) \P_1(\phi) a'_1(\phi))^2}{\kappa_1 \P_1(\phi) +
\kappa_2 \P_2(\phi)}   \label{eq:47}
\EEA

From the latter expression it follows that $\half ((-\L_{sym})^{-1} \L^{\dagger} a, \L^{\dagger} a)_{\P}=+\infty$ if
$c\not=0$. Namely, suppose adding $\frac{c}{v_1 \P_1}$ to $a_1$;
we get successively $\frac{c}{v_1\P_1}(\phi)\sim_{\phi\to\phi_1^+}
 c_1(\phi_1^+-\phi)^{-g}$; $(\frac{c}{v_1\P_1}(\phi))'\sim_{\phi\to\phi_1^+}
 c_2(\phi_1^+-\phi)^{-g-1}$; $(v_1(\phi)\P_1(\phi)a'_1(\phi))^2
 \sim _{\phi\to\phi_1^+} c_3 (\phi_1^+-\phi)^{-2}$; 
 $(\kappa_1 \P_1(\phi)+\kappa_2 \P_2(\phi))^{-1}  \sim _{\phi\to\phi_1^+} c_4 (\phi_1^+-\phi)^{1-g}$. Thus the integrand in
 (\ref{eq:47}) is not integrable in a neighborhood of $\phi_1^+$ if $c\not=0$.

\Bigskip {\em Growth functional.} By definition,
\BEQ f_{growth}(\phi)= \left(\begin{array}{c}
f_{growth,1}(\phi) \\ f_{growth,2}(\phi) \end{array}
\right) = \left(\begin{array}{c} k(1|1)\phi+k(2|1)(1-\phi) \\
k(1|2)\phi+k(2|2)(1-\phi) \end{array}\right)
\label{eq:fgrowth22}
\EEQ
and, if $a_1=a_2$,
\BEA && (f_{growth}-\lambda,a)_{\P}=\int_{\phi_2^+}^{\phi_1^+} d\phi\,  a_1(\phi)\,  \Big\{\P_1(\phi)
\ \Big( (k(1|1)\phi+k(2|1)(1-\phi)-\lambda\Big) \nonumber\\
&&\qquad + \P_2(\phi)
\ \Big(k(1|2)\phi+k(2|2)(1-\phi)-\lambda\Big) \Big\}
\EEA
As in Theorem \ref{th:main1}, we let 
$\del f_{growth}=f_{growth}-\lambda$, where $\lambda$ is the
average growth rate. We introduce the convenient bracket notation
$\langle f,g\rangle(\phi):=\sum_{s=1,2} f_s(\phi)g_s(\phi)$
for functions $f=(f_s(\phi))_{s\in {\cal S}},\  g=(g_s(\phi))_{s\in {\cal S}}$. 

\Bigskip {\em Solving for extremum.}  We find the supremum
of the concave functional of Theorem \ref{th:main2} by solving a Euler-Lagrange equation in $a_1$; namely, 
the two terms 
\BEA && \frac{\del}{\del a_1}(\del f_{growth},a)_{\P} = 
\langle \P, \del f_{growth}\rangle =
\P_1 (f_{growth,1}-\lambda) + \P_2 (f_{growth,2}-\lambda)  \label{eq:da1}
\EEA

and

\BEA && \frac{\del}{\del a_1}  \Big( \half ((-\L_{sym})^{-1} \L^{\dagger} a, \L^{\dagger} a)_{\P} \Big) = - 2 \Big( \frac{(v_1\P_1)^2}{\kappa_1 \P_1+\kappa_2 \P_2}
\, a'_1 \Big)'  \label{eq:da2}
\EEA
coming from (\ref{eq:th2}) must be equal. 
Postulating equality of the right-hand sides of  (\ref{eq:da1}) and (\ref{eq:da2})
yields $a_1$ by explicit integration. Formally,
by direct computation from (\ref{eq:th2}),
$f={\mathbb K}a$, with ${\mathbb K}=\L(-\L_{sym})^{-1}\L^{\dagger}$,
so that $\half ((-\L)^{-1}f,f)_{\P} = \half ({\mathbb K}a,a) = \half (f,a)_{\P}$.  Instead of this  computation based on
variational calculus, we can simply check, assuming equality
of (\ref{eq:da1}) and (\ref{eq:da2}), and using (\ref{eq:47}),  that
\BEA (\del f_{growth},a)_{\P}-\half ((-\L_{sym})^{-1} \L^{\dagger}a,\L^{\dagger}a)_{\P}  &=& \int a_1    \langle \P,\del f_{growth}\rangle  - \int \frac{(v_1 \P_1
a'_1)^2}{\kappa_1 \P_1+\kappa_2 \P_2} \nonumber\\
&=& - 2 \int a_1 \Big( \frac{(v_1\P_1)^2 a'_1}{\kappa_1 \P_1
+\kappa_2 \P_2}\Big)' -\int \frac{(v_1 \P_1
a'_1)^2}{\kappa_1 \P_1+\kappa_2 \P_2} \nonumber\\
&=& \int \frac{ (v_1 \P_1 a'_1)^2}{\kappa_1 \P_1 + \kappa_2 \P_2}
\EEA
(by integration by parts). 
Letting 
 $q:=\frac{(v_1\P_1)^2}{\kappa_1 \P_1+\kappa_2 \P_2}$,
  equality of (\ref{eq:da1}) and (\ref{eq:da2}) 
 is tantamount to  $\langle \P,\del f_{growth}\rangle  = -2(qa'_1)'$, whence
\BEQ \half ((-\L)^{-1} \del f_{growth},\del f_{growth}) =\int d\phi\,  q(\phi) (a'_1(\phi))^2 = 
 \frac{1}{4}\int d\phi\,  q^{-1}(\phi) \ \Big(
  \int_{\phi_2^+}^{\phi} d\phi'\, \langle \P,\del f_{growth}\rangle(\phi')
  \Big)^2
\EEQ
since $q(\phi_2^+)=0$.
 Finally, multiplying
by 4 (compare Theorem \ref{th:main1} to Theorem \ref{th:main2}),  we get our
explicit formula for the variance, in terms of the parameters
$(\pi_x)_{x=1,2},(\kappa_s)_{s=1,2}, (k(x|s))_{x,s=1,2}$ and
of the Hufton-Lin-Galla-McKane stationary distribution (\ref{eq:phi1phi2}),  (\ref{eq:2-states-P1}), (\ref{eq:2-states-P2})  only:

\begin{Theorem}[asymptotic growth-rate variance
for two states]  \label{th:final}
 Let
\BEQ \lambda = \sum_{s=1,2} \int_{\phi_2^+}^{\phi_1^+} 
d\phi\, \P_s(\phi) \del f_{growth}(\phi;s)
\EEQ
be the average growth rate (see (\ref{eq:fgrowth22})),   $I(\phi):=\int_{-\phi_2^+}^{\phi} 
d\phi'\, (\P_1 \del f_{growth,1}+\P_2 \del f_{growth,2})(\phi')$,
and $q(\phi):=\frac{(v_1\P_1)^2(\phi)}{\kappa_1 \P_1(\phi)+\kappa_2 \P_2(\phi)}$ (see (\ref{eq:v-2-states})).
Then the asymptotic  growth-rate variance (\ref{eq:varlambda})
is 
\BEQ {\mathrm{var}}(\lambda)= \int_{\phi_2^+}^{\phi_1^+}
d\phi\,  q^{-1}(\phi)\,  I^2(\phi). \label{eq:final-formula}
\EEQ
\end{Theorem}


\section{Conclusion}


We have presented in this work a derivation of the
variance of the growth rate of a general class of
mutation-selection models. Our general formula,
Theorem \ref{th:main2}, is based on the 
piecewise deterministic Markov process (PDMP) reformulation 
of the model used by Hufton-Lin-Galla-McKane. The analytic formula found by these authors
for the stationary measure of the PDMP in the simplest non-trivial 
case (two phenotypes, two environments)  makes it possible
to derive  an analytic formula both for  the average
growth-rate $\lambda$ (which was done previously   by 
Hufton et al.), and  then for its variance var$(\lambda)$ (see
our Theorem \ref{th:final}), as 
a consequence of Theorem \ref{th:main2}.

\Medskip The particular case $|{\cal X}|=|{\cal S}|=2$ already exhibits
many interesting features studied in \cite{HufLin2}, notably,
the nature of the 'optimal' phenotypic switching strategy (as characterized
by the mutation rates $\pi_1,\pi_2$), i.e. that leading to the 
largest growth-rate. It has been found that a 'bet-hedging'
strategy (with $\pi_1,\pi_2\not=0$) is favored when environmental
switching rates $\kappa_1,\kappa_2$ are not too large, whereas
a homogeneous phase, characterized by the presence of
a single phenotype, is favored in the contrary case. 

\Medskip In the companion preprint  \cite{DinLacUnt2}, 
following the strategy of \cite{DinLacUnt1}, we optimize instead of $\lambda$ alone various positive linear combinations
of $\lambda$ and of $-\var(\lambda)$. It will be interesting
to see how this changes the previous observations, in particular,
the phase diagrams.

\Medskip It seems impossible to obtain analytic formulas
for the stationary measure of the PDMP beyond the case
$|{\cal X}|=|{\cal S}|=2$, hence Theorem \ref{th:final} 
cannot be extended. However, our rather abstract variational
formula, Theorem \ref{th:main2}, which depends on the unknown
measure $\P$, allows a priori various lower bounds for 
the variance. One may hope to obtain from it uncertainty 
relations such as that obtained in \cite{GHP}, yielding a lower
bound for the variance in terms of the (squared) growth-rate
and some $\P$-dependent function playing the r\^ole of an
entropy dissipation. However, we have not been able to
prove such a formula for the moment.




\begin{thebibliography}{100}

\bibitem{Bal} Balaban N. Q., Merrin J.,  Chait R.,  Kowalik L., 
Leibler S. (2004). {\em Bacterial persistence as a phenotypic switch},  Science {\bf 305}, 1622.






\bibitem{CazHar}  Cazenave T.,  Harau A. (1998). {\em An introduction
to semilinear evolution equations}, Oxford Lecture Series in Mathematics and Its Applications {\bf 13}, Oxford Science Publications.






\bibitem{Cho} Choi P. J.,  Sai L.,  Frieda K.,  Sunney Xie X.
 (2008). {\em A stochastic single-molecule event triggers phenotype switching of a
bacterial cell},  Science {\bf 322}, 442-446.


\bibitem{DekAlo}  Dekel E.,  Alon U. (2005). {\em Optimality and evolutionary tuning of the
expression level of a protein}, Nature {\bf 436}, 588--592.

\bibitem{DinLacUnt1}  Dinis L., Lacoste D.,  Unterberger J. 
(2020). 
{\em Phase transitions in optimal strategies for gambling}, Europhysics Letters 
{\bf 131 (6)}, 60005.


\bibitem{DinLacUnt2}  Dinis L.,  Lacoste D.,  Unterberger J. 
{\em Pareto-optimal trade-off for phenotypic switching of populations in a stochastic
environment}, bioRxiv preprint available on https://www.biorxiv.org/content/10.1101/2022.01.18.476793v1.

\bibitem{EldElo}  Eldar A.,  Elowitz M. B. (2010). {\em Functional roles for noise in genetic circuits}, Nature {\bf 467}, 167--173.

\bibitem{GHP}   Gingrich T.,  Horowitz J.,  Perunov N.,  England
 J. (2016). {\em Dissipation bounds all steady-state current 
fluctuations}, Phys. Rev. Lett. {\bf 116 (12)}, 120601.

\bibitem{HufLin1}   Hufton P.,  Lin Y.,  Galla T.,  McKane
 A. (2016). {\em Intrinsic noise in systems with switching environments,} Phys. Rev. E {\bf 93 (5)}, 052119.


\bibitem{HufLin2}   Hufton P.,  Lin Y.,  Galla T.
 (2018). {\em Phenotypic switching of populations of cells in a stochastic environment},
J. Stat. Mech. {\bf 2018 (2)}, 23501.




\bibitem{KL}    Kussell E.,  Leibler S. (2005). {\em Phenotypic Diversity, Population
Growth, and Information in
Fluctuating Environments}, 
Science {\bf 309}, 2075-.




\bibitem{Lev}  Levine H.,  Jolly M. K.,  Kulkarni P. and  Nanjundiah V., editors (2020). {\em Phenotypic Switching: Implications in Biology and Medicine}, Academic Press.




\bibitem{RevYor} Revuz D.,  Yor M. (1999). {\em Continuous martingales and Brownian motion}, Springer.


\bibitem{RivWal} Mayer A., Mora T.,  Rivoire O.,  Walczak A.
 (2016). {\em Diversity of immune strategies explained by
adaptation to pathogen statistics},  Proc. Nat. Acad. Sci. {\bf 113 (31)}, 8630--8635.


\bibitem{Sha} Sharma S.,  Lee  D. Y. et al. (2010). {\em A Chromatin-Mediated
Reversible Drug-Tolerant State
in Cancer Cell Subpopulations},  Cell {\bf 141}, 69-80.

\bibitem{SkaKus} Skanata A.,  Kussell E. (2016). {\em Evolutionary Phase Transitions in Random Environments}, Phys. Rev. Lett.
{\bf  117 (3)}, 038104.



\end{thebibliography}
\end{document}